\documentclass[a4paper]{article}

\usepackage{INTERSPEECH2016}

\usepackage{graphicx}
\usepackage{amssymb,amsmath,bm}
\usepackage{textcomp}
\usepackage[acronym,nomain]{glossaries}
\usepackage{url}

\usepackage{audioicons}
\usepackage{import}
\usepackage{tikz}
\usepackage{subfigure}
\usepackage{amsfonts}
\usepackage{multirow}

\definecolor{activecolor}{RGB}{150, 150, 150}
\definecolor{loudspeakercircle}{RGB}{236, 236, 236}
\definecolor{focuscircle}{RGB}{254, 204, 0}

\newacronym{ASA}{ASA}{auditory scene analysis}
\newacronym{DNN}{DNN}{deep neural network}
\newacronym{ERB}{ERB}{equivalent rectangular bandwidth}
\newacronym{IHC}{IHC}{inner hair cell}
\newacronym{ILD}{ILD}{interaural level difference}
\newacronym{ITD}{ITD}{interaural time difference}
\newacronym{GMM}{GMM}{Gaussian mixture model}
\newacronym{EM}{EM}{expectation maximization}
\newacronym{BSS}{BSS}{blind source separation}
\newacronym{ML}{ML}{maximum likelihood}
\newacronym{TF}{TF}{time-frequency}
\newacronym{PDF}{PDF}{probability density function}
\newacronym{KEMAR}{KEMAR}{Knowles Electronics Manikin for Acoustic Research}

\newacronym{HRTF}{HRTF}{head related transfer function}
\newacronym{HRIR}{HRIR}{head related impulse responses}
\newacronym{IIR}{IIR}{Infinite Impulse Response}
\newacronym{WDS}{WDS}{Wrapped Dynamical System}
\newacronym{PS}{PS}{Periodic Scanning}
\newacronym{SPM}{SPM}{Smooth Posterior Mean}
\newacronym{UKF}{UKF}{Unscented Kalman Filter}
\newacronym{RMSE}{RMSE}{root mean square error}

\ninept

\title{An Active Machine Hearing System for Auditory Stream Segregation}
\makeatletter
\def\name#1{\gdef\@name{#1\\}}
\makeatother \name{{\em Christopher Schymura, Thomas Walther, Dorothea Kolossa}}

\address{Institute of Communication Acoustics, Ruhr-Universit\"at Bochum, Germany \\
	{\small \tt \{christopher.schymura, thomas.walther, dorothea.kolossa\}@rub.de}
}
\begin{document}
  \maketitle
  \begin{abstract}
  	This study describes a binaural machine hearing system that is capable of performing auditory stream segregation in scenarios where multiple sound sources are present. The process of stream segregation refers to the capability of human listeners to group acoustic signals into sets of distinct auditory streams, corresponding to individual sound sources. The proposed computational framework mimics this ability via a probabilistic clustering scheme for joint localization and segregation. This scheme is based on mixtures of von Mises distributions to model the angular positions of the sound sources surrounding the listener. The distribution parameters are estimated using block-wise processing of auditory cues extracted from binaural signals. Additionally, the proposed system can conduct rotational head movements to improve localization and stream segregation performance. Evaluation of the system is conducted in scenarios containing multiple simultaneously active speech and non-speech sounds placed at different positions relative to the listener. 
  \end{abstract}
  \section{Introduction}
  Human listeners have a remarkable ability to assess complex auditory scenes, even in adverse acoustic conditions. This phenomenon has been thoroughly investigated in the context of \gls{ASA}, a term that was coined by Bregman~\cite{Bregman1990}. Two essential aspects of \gls{ASA} are sound source localization and auditory stream segregation, denoting the ability of humans to build a perceptual representation of \textit{where} sounds originate from and \textit{what} kinds of sounds are present in an auditory scene \cite{Bregman1990, Blauert1999, Wang2006}. Human listeners are able to perform this task naturally in a wide range of acoustic conditions, whereas mimicking this capability by computational means still remains very challenging \cite{Wang2006}. A wide range of machine hearing systems applied to sound localization and segregation rely on a purely bottom-up information flow, see e.g.~\cite{Mandel2010, Deleforge2013}. However, various psychophysical studies have shown that the human auditory system incorporates additional top-down knowledge while assessing auditory scenes \cite{Blauert1999}. This has lead to some recent developments in the field of machine hearing, where various models that integrate top-down feedback into their processing path have been proposed \cite{Ma2015a, Ma2015b, Schymura2015}.
  
  Top-down feedback in machine hearing systems has especially been exploited in the context of binaural localization. Recent studies focus on statistical models~\cite{Ma2015a}, deterministic approximations of \glspl{HRTF}~\cite{Schymura2015} and \glspl{DNN} \cite{Ma2015b} to map \glspl{ITD} and \glspl{ILD} to the corresponding angular sound directions. By incorporating head movements, these models achieve superior localization performance compared to the static case. This arises from their ability to resolve front-back ambiguities, which are likely to occur if sound sources are positioned within the cone of confusion \cite{Blauert1999}. The results coincide with the psychophysical findings of Wallach~\cite{Wallach1940}, which indicate that human listeners also rely on head movements to improve the assessment of sound directions.
  
  This study describes a computational framework which incorporates top-down feedback into the process of auditory stream segregation. Auditory stream segregation is closely related to \gls{BSS} techniques from the field of digital signal processing. The task is to segregate binaural signals into distinct \emph{auditory streams}, where each stream corresponds to a specific sound source that is present in the acoustic scene. In contrast to conventional computational models for \gls{BSS} (see e.g.~\cite{Hoffmann2007, Rahbar2005}), the system introduced in this study uses an auditory model \cite{May2011, Decorsiere2015} for monaural and binaural feature extraction and performs a \emph{dynamic} and active assessment of the auditory scene by incorporating rotational head movements. The segregation process is based on a binaural feature analysis, which yields \emph{soft masks}, similar to many conventional \gls{BSS} methods. The soft masks are subsequently applied to extracted monaural features which are eventually used to classify the identities of all sound sources present in the scene.
  \section{System description}
  The machine hearing system described in this work is composed of different building blocks, dividing the overall process into a sequence of feature extraction, localization, auditory stream segregation, source type identification and feedback initiation. The remainder of this section gives a detailed description of each individual building block.
  \subsection{Auditory feature extraction}
  \label{subsec:feature_extraction}
  An auditory front-end as proposed in \cite{May2011} is used to extract monaural and binaural features from binaural ear signals, sampled with a rate of \(f_{\mathrm{s}} = 44.1\,\mathrm{kHz}\). Each channel of the ear signals is decomposed into \(L = 64\) auditory channels using a phase compensated gammatone filterbank. The filter center frequencies are equally distributed on the \gls{ERB} scale between \(80\,\mathrm{Hz}\) and \(8\,\mathrm{kHz}\)~\cite{Wang2006}. Half-wave rectification and low-pass filtering is applied to each frequency channel to approximate the behavior of the \glspl{IHC} \cite{Dau1996}. Subsequently, monaural and binaural features are extracted using non-overlapping, rectangularly windowed time frames with a length of \(20\,\mathrm{ms}\).
  
  The localization and segregation stage used in this study is based on two primary binaural cues, namely \glspl{ITD} and \glspl{ILD}. The \gls{ITD} between the left and the right ear signal, denoted as \(\tau_{kl}\), is estimated for each time frame \(k\) and frequency channel \(l\) by finding the time lag that corresponds to the maximum of the interaural cross-correlation function. \glspl{ILD} are estimated analogously by comparing the frame-based energy of the left and right ear \gls{IHC} signals. They are denoted as \(\delta_{kl}\) and expressed in dB. \gls{ITD} and \gls{ILD} features are combined into a 2-dimensional binaural feature vectors \(\boldsymbol{o}_{kl}~=~\begin{bmatrix}\tau_{kl}\quad\delta_{kl}\end{bmatrix}^{T}\) for each time frame and frequency channel.
  
  Monaural ratemap features \(r_{kl}\) are used to model the spectral characteristics of different sound types. They encode a spectro-temporal representation of the auditory-nerve firing rate~\cite{Brown1994}. Ratemaps are computed individually for each ear signal and frequency channel by smoothing the \gls{IHC} signal representation with a leaky integrator that has a time constant of \(8\,\mathrm{ms}\). The smoothed \gls{IHC} signal is averaged across all samples within each frame. As the stream segregation process used in this study produces soft masks, corresponding to individual weights for each time-frequency unit, these can directly be applied to the extracted ratemaps. The weighted ratemaps are subsequently used to derive spectral features for each auditory stream, which correlate to perceptual attributes of the corresponding sound source. In this study, 7 attributes are used to classify the identity of the sound source from each stream: spectral centroid~\cite{Tzanetakis2002}, spectral spread~\cite{Peeters2011}, spectral skewness~\cite{Lerch2012}, spectral kurtosis~\cite{Lerch2012}, spectral flatness~\cite{Peeters2011}, spectral crest~\cite{Peeters2011} and spectral entropy \cite{Misra2004}. These attributes are computed for each time frame, yielding the monaural feature vector \(\boldsymbol{x}_{k}\). The choice of these features was motivated by their robustness against changes in sound level, in contrast to the level-dependent ratemap features \cite{Peeters2011}.
  
  The proposed machine hearing system utilizes a block-based processing scheme to derive statistics of the spatial sound source characteristics and perform classification of individual auditory streams. Hence, a fixed number of frames \(K\) is used to compose non-overlapping \emph{blocks} of auditory features. Throughout this study, \(K = 25\) is used which corresponds to a block length of \(0.5\,\mathrm{s}\).
  \subsection{Localization framework}
  The localization framework has to meet two basic requirements: the ability to estimate the angular direction of a sound source in the entire horizontal plane and the possibility to generate a probabilistic output. Both requirements are necessary for the source segregation framework that will be described below. The localization model proposed in~\cite{Ma2015a, May2011} meets both requirements and is thus adopted for this study.
  
  The required mapping from binaural features to azimuth angles is achieved via a statistical framework based on unimodal Gaussian distributions. Individual sets of Gaussian \glspl{PDF} are considered for each frequency channel over a discrete set of equidistant azimuth angles \(\boldsymbol{\phi} = [-\pi,\, \pi)\) covering the whole horizontal plane. A set of \(M = 360\) Gaussian \glspl{PDF} per frequency channel is used here, which corresponds to an angular increment of \(1^\circ\). Hence, the localization model is composed of \(360 \times 64\) Gaussian \glspl{PDF}, where each \gls{PDF} is specified by its mean vector and covariance matrix, represented as a set of model parameters \(\theta_{l}^{(m)},\,m=1\,\ldots,\,M\). The model parameters are estimated using the 2-dimensional binaural features as described in Sec.~\ref{subsec:feature_extraction}. The training features are computed using anechoic \gls{HRIR} of the \gls{KEMAR} dummy head \cite{Wierstorf2011} with white noise stimulus signals. Full covariance matrices were assumed for all models during training.
  
  \noindent Evaluating the posterior probability
  \begin{equation*}
  p(\phi^{(m)}\,|\,\boldsymbol{o}_{kl}) \propto \frac{p(\boldsymbol{o}_{kl}\,|\,\theta_{l}^{(m)})}{\sum_{i = 1}^{M} p(\boldsymbol{o}_{kl}\,|\,\theta_{l}^{(i)})}
  \end{equation*}
  for each discrete angular position \(\phi^{(m)}\) yields an estimate of the azimuthal sound source direction, computed as the \gls{ML} solution
  \begin{equation}
  \label{eqn:relative_azimuth}
  \tilde{\phi}_{kl} = \underset{\phi^{(m)}}{\arg \max}~p(\phi^{(m)}\,|\,\boldsymbol{o}_{kl})
  \end{equation}
  for each \gls{TF} unit. As Eq.~\eqref{eqn:relative_azimuth} produces an estimate of the source azimuth \emph{relative} to the listeners look direction \(\psi_{k}\), the latter has to be taken into account explicitly. Therefore, the absolute angular direction of the sound source is computed as
  \begin{equation}
  \label{eqn:absolute_azimuth}
  \phi_{kl} = \Big((\tilde{\phi}_{kl} + \psi_{k} + \pi)\,\mathrm{mod}\,{2\pi}\Big) - \pi
  \end{equation}
  for each \gls{TF} unit. For a given set of estimated source positions in a single signal block \(\{\phi_{kl}\},~k=1,\,\ldots,\,K,~l=1,\,\ldots,\,L\), a vector of observations is derived by changing the double element index \(kl\) to a single index \(n=1,\,\ldots,\,N\) with \(N=K \cdot L\) over all time and frequency units and stacking the individual observations according to
  \begin{align}
  \boldsymbol{\phi} &= \begin{bmatrix}
  \phi_{11},\,\ldots,\,\phi_{K1},\,\phi_{12},\,\ldots,\,\phi_{K2},\ldots,\,\phi_{KL}
  \end{bmatrix}^{T}, \notag \\
  &= \begin{bmatrix}
  \phi_{1},\,\ldots,\,\phi_{N}
  \end{bmatrix}^{T}. \label{eqn:loc_observations}
  \end{align}
  \subsection{Auditory stream segregation}
  The auditory stream segregation framework assigns estimates of angular positions to each time-frequency unit of an acquired signal block. This representation serves as the basis for a subsequent clustering step. However, conventional clustering techniques like \(k\)-means \cite{Macqueen1967} or \glspl{GMM} \cite{Dempster1977} might not be suitable for the problem at hand, since the available observations are azimuth angles, originating from a circular probability distribution bounded in \([-\pi,\,\pi]\). Therefore, an alternative clustering technique is applied here, which is based on a mixture of von Mises distributions \cite{Banerjee2005}. Similar approaches have already been proposed in the context of sound source localization and tracking \cite{Traa2014, Markovic2012}. In this work, it is applied in the context of auditory stream segregation, by using dummy head \glspl{HRIR} with a binaural sensor and an auditory feature extraction stage. The \gls{PDF} of a von Mises distribution is defined as
  \begin{equation}
  \label{eqn:von_mises}
  \mathcal V \mathcal M(\phi,\,|\,\mu,\,\kappa) = \frac{1}{2\pi I_{0}(\kappa)} \exp \Big\{\kappa \cos (\phi - \mu)\Big\},
  \end{equation}
  where \(\phi \in [-\pi,\,\pi]\) is an angle, \(\mu\) is the circular mean, \(\kappa\) is the concentration parameter and \(I_{i}(\cdot)\) is the modified \(i\)-th order Bessel function. Hence, the \gls{PDF} of a mixture of von Mises distributions can be derived from Eq.~\eqref{eqn:von_mises} as
  \begin{equation}
  \label{eqn:von_mises_mixture}
  p(\phi\,|\,\boldsymbol{\pi},\,\boldsymbol{\mu},\,\boldsymbol{\kappa}) = \sum_{c = 1}^{C} \pi_{c} \mathcal V \mathcal M(\phi\,|\,\mu_{c},\,\kappa_{c}),
  \end{equation}
  where \(\boldsymbol{\pi} = [\pi_{1},\,\ldots,\,\pi_{C}]^{T}\) are the mixture weights satisfying \(\sum_{c=1}^{C} \pi_{c} = 1\), \(\boldsymbol{\mu} = [\mu_{1},\,\ldots,\,\mu_{C}]^{T}\) denote the circular means and \(\boldsymbol{\kappa} = [\kappa_{1},\,\ldots,\,\kappa_{C}]^{T}\) are the concentration parameters corresponding to each of the \(C\) mixture components.
  
  The stream segregation process applied in this study is based on the assumption that the number of active sound sources is known \emph{a priori}. Therefore, the number of mixture components \(C\) is fixed according to this prior knowledge. For a given set of estimated source positions, the log-likelihood of the \gls{PDF} introduced in Eq.~\eqref{eqn:von_mises_mixture} can be expressed as
  \begin{equation}
  \label{eqn:von_mises_loglik}
  \mathcal L(\boldsymbol{\phi}\,|\,\boldsymbol{\pi},\,\boldsymbol{\mu},\,\boldsymbol{\kappa}) = \sum_{n=1}^{N} \log \Big(\sum_{c=1}^{C}\pi_{c}\mathcal V \mathcal M(\phi_{n}\,|\,\mu_{c},\,\kappa_{c})\Big).
  \end{equation}
  The parameters of Eq.~\eqref{eqn:von_mises_loglik} are estimated using an \gls{EM} scheme based on the approach presented in~\cite{Hung2012}. The parameter estimates at each maximization step are given as
  \begin{align}
  \mu_{c} &= \mathrm{atan2} \Big(\sum_{n=1}^{N} \gamma_{nc} \sin (\phi_{n}),\, \sum_{n=1}^{N} \gamma_{nc} \cos (\phi_{n})\Big), \label{eqn:mstep_mu} \\
  \kappa_{c} &= A^{-1}\Big(\frac{\sum_{n=1}^{N} \gamma_{nc} \cos(\phi_{n} - \mu_{c})}{\sum_{n=1}^{N} \gamma_{nc}}\Big), \label{eqn:mstep_kappa} \\
  \pi_{c} &= \frac{1}{N} \sum_{n=1}^{N} \gamma_{nc}, \label{eqn:mstep_pi}
  \end{align}
  with
  \begin{equation}
  A(x) = \frac{I_{1}(x)}{I_{0}(x)}. \label{eqn:inverse_kappa}
  \end{equation}
  and responsibilities
  \begin{equation}
  \gamma_{nc} = \frac{\pi_{c} \mathcal V \mathcal M(\phi_{n}\,|\,\mu_{c},\,\kappa_{c})}{\sum_{j=1}^{C} \pi_{j} \mathcal V \mathcal M(\phi_{n}\,|\,\mu_{j},\,\kappa_{j})} \label{eqn:von_mises_gamma}
  \end{equation}
  computed during the E-step. Estimating the concentration parameters \(\kappa_{c}\) requires inverting the function given in Eq.~\eqref{eqn:inverse_kappa}. This problem cannot be solved analytically, therefore the inverse function has to be approximated. In this study, the approximation scheme introduced in~\cite{Best1981} is applied to estimate the concentration parameters. The \gls{EM} algorithm utilises Eqs.~\eqref{eqn:mstep_mu}--\eqref{eqn:inverse_kappa} to incrementally update the parameter estimates during the optimization process. The initial model parameters are computed using the circular \(k\)-means algorithm described in \cite{Banerjee2005}.
  
  Following the parameter estimation procedure, the model described in Eq.~\eqref{eqn:von_mises_mixture} is used to derive soft masks for all active sound sources, denoted by \(c=1,\,\ldots,\,C\). The masking coefficients \(\beta_{kl}^{(c)}\) for each \gls{TF}-unit are computed by evaluating the normalized likelihood of the corresponding mixture component, given the estimated azimuth angle:
  \begin{equation}
  \label{eqn:soft_masks}
  \beta_{kl}^{(c)} = \frac{\mathcal{VM}(\phi_{kl}\,|\,\mu_{c},\,\kappa_{c})}{\sum_{j=1}^{C} \mathcal{VM}(\phi_{kl}\,|\,\mu_{j},\,\kappa_{j})}
  \end{equation}
  Note, that the soft-mask estimation assumes a uniform prior over individual mixture components, hence the mixture weights \(\pi_{c}\) are discarded in Eq.~\eqref{eqn:soft_masks}. The estimated soft masks are subsequently applied to the extracted ratemap features, yielding an auditory stream \(\tilde{r}_{kl}^{(c)}~=~r_{kl}~\cdot~\beta_{kl}^{(c)}\) for each source. Individual sets of perceptual attributes \(\tilde{\boldsymbol{x}}_{k}^{(c)}\) are then derived from these auditory streams as described in Sec.~\ref{subsec:feature_extraction}.
  \subsection{Source classification}
  \label{subsec:classification}
  The identities of the sound sources present in the auditory scene are inferred by a classification scheme based on \glspl{GMM}. Let \(\lambda_{s}\) represent the characteristics of a sound source in terms of the monaural auditory features described in Sec.~\ref{subsec:feature_extraction}. Given a set of source models \(s = 1,\,\ldots,\,\mathcal S\) and a vector of perceptual attributes \(\tilde{\boldsymbol{x}}_{k}^{(c)}\), the posterior probability of source model \(s\) at frame \(k\) can be computed as
  \begin{equation}
  \label{eqn:post_prob_clf}
  p(\lambda_{s}\,|\,\tilde{\boldsymbol{x}}_{k}^{(c)}) = \frac{p(\tilde{\boldsymbol{x}}_{k}^{(c)}\,|\,\lambda_{s})\,p(\lambda_{s})}{\sum_{s} p(\tilde{\boldsymbol{x}}_{k}^{(c)}\,|\,\lambda_{s})\,p(\lambda_{s})}.
  \end{equation}
  A uniform prior \(p(\lambda_{s})\) is assumed in this study. As the auditory stream segregation framework described in the previous section utilizes a block-based processing scheme, the posterior probability in Eq.~\eqref{eqn:post_prob_clf} has to be extended accordingly. This is achieved by averaging the frame posteriors across time to produce a posterior probability of source model \(s\) given a set of perceptual attributes derived from a block of \(K\) frames. The source identity of a block within the \(c\)-th auditory stream is the considered to be the source model \(\hat{s}\) that maximizes the posterior probability according to
  \begin{equation*}
  \hat{s} = \underset{s}{\arg \max} \frac{1}{K}\sum_{k = 1}^{K} p(\lambda_{s}\,|\,\tilde{\boldsymbol{x}}_{k}^{(c)}).
  \end{equation*}
  \subsection{Feedback through rotational head movements}
  Two different head rotation schemes and their effect on localization and segregation performance are investigated in this study. They are partially adopted from previous work presented in \cite{Schymura2015}. The case when no rotational head movements are applied serves as a baseline for the performance comparison. The first approach performs random head movements based on the dynamics equation
  \begin{equation*}
  \psi_{k} = \psi_{k-1} + \frac{\pi}{180}u_{k},\quad u_{k} \sim \mathcal N(0, 1).
  \end{equation*}
  This is a purely feed-forward control strategy, as it uses no information provided by the auditory stream segregation stage.
  
  In contrast, the second approach investigated in this study imposes a feedback loop by turning the head towards the most uncertain of the estimated source positions. The concentration parameters of the mixture of von Mises distributions given in Eq.~\eqref{eqn:von_mises_mixture} are used as a measure of uncertainty in this approach, yielding the index of the most uncertain source position as
  \begin{equation*}
  \tilde{c} = \underset{c}{\arg \min}~\kappa_{c}.
  \end{equation*}
  Feedback is subsequently initiated using the dynamics equation
  \begin{equation}
  \label{eqn:feedback}
  \psi_{k} = \psi_{k-1} + \alpha (\mu_{\tilde{c}} - \psi_{k - 1}),
  \end{equation}
  where \(\alpha\) is a fixed gain factor set to \(\alpha = 5\) throughout all experiments. For both proposed approaches, possible head rotations were restricted to the range \([-80^\circ,\,80^\circ]\) relative to the initial look direction which was fixed at \(90^\circ\).
  \section{Evaluation}
  \subsection{Sound database}
  A collection of speech and non-speech sounds was created using the GRID corpus \cite{Cooke06} and a publicly available sound database\footnote{\url{https://www.freesound.org/}}. Five different classes of sounds with diverse spectro-temporal complexity were selected: \emph{female speech}, \emph{alarm/siren}, \emph{dog barking}, \emph{car engine} and \emph{piano}. Silence periods in all sound files were manually labeled.
  \subsection{Experimental setup}
  \begin{figure}
  	\centering
  	\tikzstyle{loudspeaker} = [basic loudspeaker, draw, fill=white
, minimum height=2mm 	    
, minimum width=1mm		    
, inner sep=0pt
, relative cone width=1	
, relative cone height=2	    
]

\tikzstyle{active} = [fill=black!50]
\tikzstyle{focus} = [draw, fill=white, circle, minimum size=2mm, inner sep=0pt]
\tikzstyle{source} = [draw, fill=black, circle, minimum size=1mm, inner sep=0pt]
\tikzstyle{head} = [fill=white]

\pgfdeclarelayer{background}
\pgfsetlayers{background,main}

\begin{tikzpicture}

\pgfmathsetmacro{\R}{2.5}
\pgfmathsetmacro{\Rlabel}{\R+0.2}

\pgfmathsetmacro{\Rhead}{0.4}
\pgfmathsetmacro{\alphahead}{40}
\pgfmathsetmacro{\Rear}{0.2*\Rhead}
\pgfmathsetmacro{\alphaear}{95}
\pgfmathsetmacro{\Rnose}{1.2*\Rhead}
\pgfmathsetmacro{\alphanose}{7}

\pgfmathsetmacro{\Rcoord}{\R*1.2}
\pgfmathsetmacro{\alpha}{90}
\pgfmathsetmacro{\alphainc}{30}

\pgfmathsetmacro{\scalefactor}{1}


\draw[thick, -latex] (-0.90*\Rcoord,0) -- (\Rcoord,0) node[at end, below] {$x$};
\draw[thick, -latex] (0,-0.50*\Rcoord) -- (0,\Rcoord) node[at end, left] {$y$};

\draw[head] (\alphahead+\alphaear:\Rhead) circle (\Rear);
\draw[head] (\alphahead-\alphaear:\Rhead) circle (\Rear);
\draw[head] (\alphahead+\alphanose:\Rhead) --
            (\alphahead:\Rnose) -- 
            (\alphahead-\alphanose:\Rhead);
\draw[head] (0,0) circle (\Rhead);

\draw[head, dashed]  (0,0) -- (\alphahead:1.05*\R);
\draw[-latex] (0:4.75*\Rhead) arc (0:\alphahead:4.75*\Rhead);
\node [left] at (0.5*\alphahead:4.75*\Rhead) {$\psi_k$};

\node[loudspeaker, rotate=150-180] (speaker) at (150:\scalefactor*2.5) {};
\node[loudspeaker, rotate=110-180] (speaker) at (110:\scalefactor*2.5) {};
\node[loudspeaker, rotate=70-180] (speaker) at (70:\scalefactor*2.5) {};
\node[loudspeaker, rotate=30-180] (speaker) at (30:\scalefactor*2.5) {};

\begin{pgfonlayer}{background}
  \fill[fill=loudspeakercircle] (0,0) -- (10:1.05*\R) arc (10:170:1.05*\R);
  \fill[fill=focuscircle] (0,0) -- (0:4.75*\Rhead) arc (0:\alphahead:4.75*\Rhead);
\end{pgfonlayer}

\end{tikzpicture}
  	\caption{Layout of the acoustic scene used during the evaluation. Sound sources can be positioned at \(30^\circ\), \(70^\circ\), \(110^\circ\) and \(150^\circ\). The gray area indicates the range of possible head rotations.}
  	\label{fig:exp_setup}	
  \end{figure}
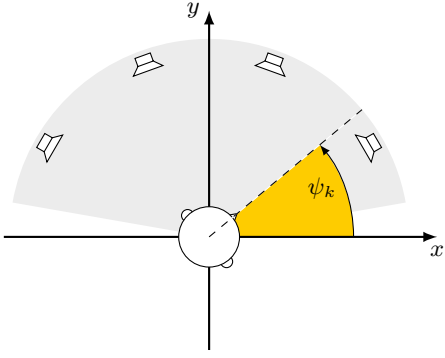
  The proposed framework was evaluated in simulated acoustic scenes, which were generated using anechoic \glspl{HRIR} \cite{Wierstorf2011}. Three different scenarios with either 2, 3 or 4 simultaneously active sound sources were considered. The sources were positioned randomly at fixed azimuth angles relative to the dummy head. An overview of the general acoustic scene configuration used for the evaluation is depicted in Fig.~\ref{fig:exp_setup}. It is worth mentioning that, even though the sources were all placed within the frontal hemisphere, the system had no prior knowledge about the azimuthal positions. This was ensured by initializing the stream segregation stage with uniform circular distributions.
  
  The acquired sound database was partitioned into 10 folds of training and test sets, allowing cross-validation of the dataset in all evaluation scenarios. Monaural feature vectors were extracted for each sound class from the training files and used to train the \gls{GMM}-based classifiers described in Sec.~\ref{subsec:classification}. All \glspl{GMM} were trained using \gls{EM} iterations with a fixed number of 16 mixture components and full covariance matrices. Silence periods were excluded from the audio files during training.
  
  During each cross-validation step, 30 simulations with a fixed duration of \(3\,\mathrm{s}\) each were conducted for testing. The specific setup of each simulation was composed of randomly chosen sounds from the test set. Localization and classification performance were measured by computing the block-wise, cumulative circular \gls{RMSE} and classification error rate, respectively.
  \subsection{Results and discussion}
  The results obtained in all evaluation scenarios are summarized in Tab.~\ref{tab:results}. The feedback strategy introduced in Eq.~\eqref{eqn:feedback} outperforms both the static case and the feed-forward control scheme based on random head movements in all evaluated scenarios. Furthermore, random head movements yield no significant improvements with respect to the baseline throughout all conducted experiments. This effect is similar to the results described in \cite{Schymura2015}, where it was shown that control strategies based on a feed-forward paradigm are not as beneficial for localization performance as closed-loop feedback control.
  
  The results obtained in this work indicate that classification performance of the proposed machine hearing system is highly dependent on localization accuracy. An increasing number of simultaneously active sound sources results in a severe drop in localization performance, which subsequently also reduces the stream segregation abilities of the system. This effect is caused by the rather basic localization framework used in this study, which is based on Gaussian \glspl{PDF}. A more sophisticated model would most certainly yield in better azimuth estimations, thus providing more accurate features to the clustering stage of the stream segregation framework.
  \begin{table}[t]
  	\caption{\label{tab:results} {\it This table summarizes the averaged localization and classification performance achieved in all conducted experiments for different head rotation strategies. The localization error is given as cumulative circular \gls{RMSE} in degrees, whereas classification error is depicted as classification error rate in percent. The best performances achieved in each of the scenarios are depicted in bold font.}}
  	\vspace{2mm}
  	\centerline
  	{
  		\begin{tabular}{| l | c | c |}
  			\hline
  			Head rotation & Localization error & Classification error \\
  			\hline \hline
  			\multicolumn{3}{|l|}{\textbf{Scenario 1:} 2 active sound sources} \\
  			\hline
  			None & 30.12 & 31.67 \\
  			Random & 28.91 & 31.07 \\
  			Feedback & \textbf{20.91} & \textbf{23.33} \\
  			\hline \hline
  			\multicolumn{3}{|l|}{\textbf{Scenario 2:} 3 active sound sources} \\
  			\hline
  			None & 64.52 & 67.75 \\
  			Random & 62.67 & 67.78 \\
  			Feedback & \textbf{61.82} & \textbf{63.19} \\
  			\hline \hline
  			\multicolumn{3}{|l|}{\textbf{Scenario 3:} 4 active sound sources} \\
  			\hline
  			None & 78.94 & 77.96 \\
  			Random & 78.97 & 78.35 \\
  			Feedback & \textbf{70.45} & \textbf{74.79} \\     			
  			\hline
  		\end{tabular}
  	}
  \end{table}
  \section{Conclusions}
  This study presented an active machine hearing system for auditory stream segregation in multi-source scenarios. The proposed system is based on a framework for joint localization and stream segregation, using a probabilistic clustering scheme to assign individual \gls{TF} units to azimuthal sound source positions. The framework is termed \emph{active}, because it is able to dynamically assess the auditory scene by conducting head movements. Experimental results have indicated, that rotational head movements based on a closed-loop feedback control scheme increase localization and stream segregation performance. This was shown in simulated acoustic environments using a combined stream segregation and classification system to infer the identities of different speech and non-speech sounds in multi-source scenarios.
  
  Future developments will focus on improving the localization framework used in this study, e.g. by incorporating \glspl{GMM} or \glspl{DNN} for predicting azimuth angles for each \gls{TF} unit. Additionally, different head rotation strategies may yield further improvements of auditory stream segregation performance. In this context, it will be especially interesting to investigate head movements conducted by human listeners in multi-source scenarios and apply according strategies to the proposed computational system.
  \section{Acknowledgements}
  This research has been supported by EU FET grant \textsc{Two!Ears}, ICT-618075.
  \newpage
  \eightpt
  \bibliographystyle{IEEEtran}
  \bibliography{mybib}
\end{document}